\documentclass[aps,pra,twocolumn,amsmath,amssymb,floatfix,showkeys,showpacs,reprint]{revtex4-1}
\usepackage{graphicx}
\usepackage{bm} 
\usepackage{color}
\usepackage{graphics} 
\usepackage{graphicx} 
\usepackage{epsfig}
\usepackage{amssymb}
\usepackage{amsmath} 
\usepackage{amsthm}
\usepackage{amstext}
\usepackage{dcolumn} 
\usepackage{makeidx}
\usepackage{subfigure}
\usepackage{float}

\begin{document}

\title{Multiphoton Raman transitions and Rabi oscillations in driven spin systems}

\author{A. P. Saiko}
\email{saiko@physics.by}
\affiliation{Scientific-Practical Material Research Centre, Belarus National Academy of Sciences, 19 P. Brovka str. , Minsk 220072 Belarus}

\author{R. Fedaruk}
\affiliation{Institute of Physics, Faculty of Mathematics and Physics, University of Szczecin, 15 Wielkopolska str. , 70-451, Szczecin, Poland}

\author{S. A. Markevich}
\affiliation{Scientific-Practical Material Research Centre, Belarus National Academy of Sciences, 19 P. Brovka str. , Minsk 220072 Belarus}

\date{\today}

\begin{abstract}
In the framework of the non-secular perturbation theory based on the Bogoliubov averaging method, the coherent dynamics of multiphoton Raman transitions in a two-level spin system driven by an amplitude-modulated microwave field is studied. Closed-form expressions for the Rabi frequencies of these transitions have been obtained beyond the rotating wave approximation for the low-frequency driving component. It is shown that spin states dressed by the high-frequency component of the driving field are shifted due to the Bloch-Siegert-like effect caused by antiresonant interactions with the strong low-frequency driving. We predict that with increasing the order of the Raman transition the Rabi frequency decreases and the contribution of the Bloch-Siegert shift to this frequency becomes dominant. It is found that the amplitude and phase of the Rabi oscillations strongly depend on the initial phase of the low-frequency field as well as on detuning from multiphoton resonance. The recent experimental data for the second- and third-order Raman transitions observed for nitrogen-vacancy center in diamond [Z. Shu, et al., arXiv:1804. 10492] are well described in the frame of our approach. Our results provide new possibilities for coherent control of quantum systems.
\end{abstract}

\pacs{42. 50. Ct, 42. 50. Dv, 42. 50. Hz, 76. 30. Mi}

\maketitle

\section{INTRODUCTION}

The coherent dynamics of two-level quantum systems (qubits) driven by electromagnetic fields is successfully used for studying and control of a wide range of physical objects including, among others, spins \cite{pp1}, atoms \cite{pp2}, artificial atoms such as quantum dots \cite{pp3} and superconducting qubits \cite{pp4}. In particular, this dynamics is extremely important for quantum information processing \cite{pp1,pp5}, quantum sensing \cite{pp6}, and the realization of new exotic phases, such as topological Floquet insulators \cite{pp7} and time crystals \cite{pp8, pp9}. Rabi oscillations are the cyclic behaviour of the probability of finding the two-level system in the excited state and represent the basic phenomenon used for coherent manipulation of quantum states. The coherent dynamics of qubits can be described in terms of dressed states \cite{pp10}. The dressing of qubit by the electromagnetic field gives rise to new energy levels of the coupled field-qubit system. The splitting of each bare level is characterized by the Rabi frequency. Stimulated transitions between the dressed states open an additional tool for coherent quantum manipulation and control \cite{pp11}. In particular, such transitions are effectively excited by the second field with the frequency closed to the Rabi frequency determining the splitting between the dressed states of the driven two-level system. This so-called Rabi resonance has been observed for spin ensembles \cite{pp12,pp13} and a single spin \cite{pp14} in EPR, NMR \cite{pp15,pp16,pp17,pp18} as well as for atoms in the optical range \cite{pp19,pp20,pp21}. Additional resonances occur at the subharmonics of the Rabi frequency \cite{pp17,pp18,pp19,pp22}. The coherent dynamics of the dressed-state transitions has been studied directly in time-resolved experiments by recording the Rabi oscillations between the dressed states \cite{pp12,pp13,pp14,pp16,pp18,pp20,pp21}. Since the strength of the driving field inducing transitions between the dressed states is often comparable with the Rabi frequency, the rotating wave approximation (RWA) is broken and the contribution of the antiresonant (non-RWA) terms to the coupling Hamiltonian must be taken into account to explain fully the experimental observations \cite{pp13,pp14,pp23}. In these papers the Rabi resonance has been observed when the first driving field was in resonance with the spin system. Aiming to illustrate the effects of the non-RWA terms of the transverse field in the strongly driven two-level system under the low-frequency modulation, the counterrotating-hybridized rotating-wave method has been developed \cite{pp29} and the fluorescence spectrum of such system has been obtained \cite{pp30}.

Recently, so-called Floquet Raman transitions have been observed in the driven solid-state spin system of nitrogen-vacancy center in diamond \cite{pp24}. The system was driven by the microwave field with its low-frequency amplitude modulation. The microwave frequency was detuned from the resonant frequency of the two-level system. Raman transitions between dressed spin states were excited by the low-frequency field when multiphoton resonances (termed also Floquet resonances \cite{pp25}) were realized. To describe the Rabi frequencies of these transitions, in the frame of Floquet theory the effective Hamiltonian for the two-level system has been found and the analytical expressions for the corresponding Rabi frequencies have been obtained in the RWA \cite{pp24}. However, the calculated Rabi frequencies of the Raman transitions were significantly smaller that the measured those. The correct values of the observed frequencies were obtained by numerical simulation. Non-resonant interactions of the low-frequency field with the dressed spin states, which were neglected in the used effective Floquet Hamiltonian, may cause the difference between the analytical estimates and the experimental data. These interactions can be significant when the low-frequency driving strength becomes comparable with the splitting between the dressed states. Such interactions shift the dressed energy levels and renormalize the interaction strength characterizing by the Rabi frequency of the Raman transition.

In the present paper, we propose a theoretical description of multiphoton Raman transitions in a two-level spin system driven by an amplitude-modulated microwave field. The description is obtained beyond the rotating wave approximation for the low-frequency
driving component. We demonstrate that the construction of the effective Hamiltonian in the framework of the non-secular perturbation theory based on the Bogoliubov averaging method allows us to take into account the non-resonant processes up to the third-order of the perturbation. The effective Hamiltonian for multiphoton Raman transitions is introduced in Sec. II. The analytical description of the dynamics of  Raman transitions is presented in Sec. III. Here the effective strength of Raman transitions is given taking into account the Bloch-Siegert-like effect. The third-order correction to the Rabi frequency is considered in the Appendix. Effects of the initial phase of the low frequency field and detuning of its frequency from multiphoton resonances are considered. We test our calculations using the recent experimental data for nitrogen-vacancy center in diamond \cite{pp24}. The obtained analytical description of multiphoton Raman transitions is in a good agreement with the experimental results and demonstrates the significant contribution of the Bloch-Siegert-like effect to the observed Rabi frequencies.

\section{THE EFFECTIVE HAMILTONIAN FOR MULTIPHOTON RAMAN TRANSITIONS}

To observe Raman transitions between Floquet dressed states of an initially two-level spin system, a microwave field described by $V(t)=\Delta _{x} \cos (\omega _{d} t)+2A\cos (\omega _{d} t)\sin (\omega t+\psi )$ is synthesized \cite{pp24}. Here $\cos (\omega _{d} t)$ describes the high-frequency component of the field, $\sin (\omega t+\psi )$ represents the low-frequency component with the initial phase $\psi $, and $\Delta _{x} $, $A$ $<$$<$$\omega _{d} $. In particular, for the NV center an effective two-level system is realized when the microwave field excites transitions between the spin sublevels ${\left| 0 \right\rangle} $ and ${\left| -1 \right\rangle} $ of this center, while the level ${\left| +1 \right\rangle} $ is far detuned. The Hamiltonian of the two-level system at such driving can be written as $H_{lab} =\frac{\Delta E}{2} \sigma ^{z} +\Delta _{x} \cos (\omega _{d} t)\sigma ^{x} +2A\cos (\omega _{d} t)\sin (\omega t+\psi )\sigma ^{x} $, where $\Delta E$ is the transition energy between the levels ${\left| 0 \right\rangle} $and ${\left| -1 \right\rangle} $; $\sigma ^{z} $and $\sigma ^{x} $ are Pauli operators. We use the frame rotating with the driving field frequency $\omega _{d} $ and the RWA for this field (since the conditions $\omega $, $\Delta _{x} $, $A < < \omega _{d} $ are fulfilled). The obtained Hamiltonian for an analysis of Raman transitions is
\begin{equation} \label{eq_1}
H=\frac{\Delta _{z} }{2} \sigma ^{z} +\frac{\Delta _{x} }{2} \sigma ^{x} +A\sin (\omega t+\psi )\sigma ^{x},
\end{equation}
where $\Delta _{z} =\Delta E-\omega _{d} $. The dynamics of the system under study is described by the Liouville equation for the density matrix $\rho $: $i\partial \rho /\partial t=H\rho $ (in the following we take $\hbar =1$). Rotating the frame around the \textit{y} axis by angle of $\theta $ ($\rho \to \rho _{1} =U_{1}^{\dagger} \rho U_{1} $, $U_{1} =e^{-i\theta \sigma ^{y} /2} $, and $\sigma ^{y} =(\sigma ^{+} -\sigma ^{-} )/i$), we obtain the same equation with the Hamiltonian $H_{1} =U_{1}^{\dagger} HU_{1} =\frac{\omega _{0} }{2} \sigma ^{z} +A\cos \theta \sin (\omega t+\psi )\sigma ^{x} +A\sin \theta \sin (\omega t+\psi )\sigma ^{z} $, where $\omega _{0} =\sqrt{\Delta _{z}^{2} +\Delta _{x}^{2} } $, $\sin \theta =\Delta _{x} /\omega _{0} $, $\cos \theta =\Delta _{z} /\omega _{0} $. After the second canonical transformation $\rho _{1} \to \rho _{2} =U_{2}^{\dagger} \rho _{1} U_{2} $ with $U_{2} =\exp \left\{-i\left[\omega _{0} t-\frac{2A\sin \theta }{\omega } \cos (\omega t+\psi )\right]\frac{\sigma ^{z} }{2} \right\}$, we obtain the Liouville equation for $\rho _{2} $ with the Hamiltonian
$$H_{2} =U_{2}^{\dagger} H_{1} U_{2} -iU_{2}^{\dagger} \frac{\partial U_{2} }{\partial t} =$$
$$\frac{A}{2i} \cos \theta \left[\sigma ^{+} \sum _{n=-\infty }^{\infty }J_{n} (a)e^{-in\pi /2}\right. $$
\begin{equation} \label{eq_2}
\Biggl. \left(e^{i(n+1)\omega t} e^{i(n+1)\psi } -e^{i(n-1)\omega t} e^{i(n-1)\psi } \right)e^{i\omega _{0} t} +h. c. \Biggr],
\end{equation}
where $J_{n} (a)$ is the Bessel function of the first kind and $a=2A\sin \theta /\omega $.

Now we consider multiphoton Raman transitions when the resonance condition $\omega _{0} /k=\omega -\delta $ is fulfilled for $k=1,2,3,...$, where detuning $\delta$ from the exact resonance is introduced and $\left|\delta \right|<<\omega $. The Hamiltonian $H_{2} $ contains an infinite sum of oscillating harmonics with the frequencies which are integer multiples of the frequency $\omega $. There are no oscillations for $n=-k+1$ and $n=-k-1$. Therefore, the terms of the sum with these \textit{n} give the largest contribution. These terms correspond to the RWA. However, the other oscillating terms can be significant, if the strong coupling condition $0. 1<A/\omega <1$ is fulfilled. The contribution of such oscillating terms can be taken into account using the Bogoliubov averaging method \cite{pp26}. This method allows us to construct in the framework of the non-secular perturbation theory some time-independent effective Hamiltonian. The averaging procedure up to the second order in $A\cos \theta /\omega $ (see \cite{pp26,pp27}) gives the following effective Hamiltonian: $H_{2} \to H_{eff} =H_{2}^{(1)} +H_{2}^{(2)} $, where
$$H_{2}^{(1)} =<H_{2} (t)>,$$
\begin{equation} \label{eq_3}
H_{2}^{(2)} =\frac{i}{2} <[\int _{}^{t}d\tau (H_{2} (\tau )-<H_{2} (\tau )>),H_{2} (t) ]>.
\end{equation}

Here the symbol $\langle . . . \rangle $ denotes time averaging over rapid oscillations of the type $\exp (\pm im\omega t)$ given by $\langle O(t)\rangle =\frac{\omega }{2\pi } \int _{0}^{{2\pi \mathord{\left/ {\vphantom {2\pi \omega }} \right. \kern-\nulldelimiterspace} \omega } }O(t)dt $. The upper limit \textit{t} of the indefinite integral indicates the variable on which the result of the integration depends, and square brackets denote the commutation operation.

Calculations based on Eq. \eqref{eq_3} give:
$$H_{2}^{(1)} (k)=(-1)^{k+1} \frac{\Omega _{k} }{2} (\sigma ^{+} e^{-ik(\psi -\pi /2)} e^{-ik\delta t} +h. c. ),$$
\begin{equation} \label{eq_4}
H_{2}^{(2)} (k)=\frac{\omega _{k}^{BS} }{2} \sigma ^{z},
\end{equation}
where
$$\Omega _{k} =2k\frac{J_{k} (a)}{a} A\cos \theta,$$
\begin{equation}
\label{eq_5}
\omega _{k}^{BS} =\frac{A^{2} \cos ^{2} \theta }{2\omega }\times \\
\end{equation}
$$\times\left\{\sum _{n\ne -k-1}\frac{J_{n}^{2} +J_{n} J_{n+2} }{n+k+1} +\sum _{n\ne -k+1}\frac{J_{n}^{2} +J_{n} J_{n-2} }{n+k-1} \right\}. $$
Here $\Omega _{k} $ is the Rabi frequency of the $k$-th order Raman transition in the RWA, $\omega _{k}^{BS} $ is the Bloch-Siegert-like frequency shift for the $k$-th order transition and all Bessel functions are evaluated at point $a$. This shift is caused by the non-resonant terms in the Hamiltonian \eqref{eq_2}. The Bessel functions $J_{n} (a)$ in Eq. \eqref{eq_5} for $\Omega _{k} $ appear due to taking into account virtual multiphoton transitions, in which the number of absorbed (emitted) photons exceeds by $\left|k\right|$ the number of emitted (absorbed) photons.

Below we consider the total effective Hamiltonian $H_{eff} (k)$ which is the sum of the Hamiltonians $H_{2}^{(1)} (k)$ and $H_{2}^{(2)} (k)$ over every possible value of $k$.

\begin{widetext}
\section{THE DYNAMICS OF RAMAN TRANSITIONS }

For the $k$-th order Raman transition the dynamics of the system under study is described by the Liouville equation for the density matrix $\rho _{3}^{(k)} $: $i\partial \rho _{3}^{(k)} /\partial t=\tilde{H}_{eff} (k)\rho _{3}^{(k)} $, where $\tilde{H}_{eff} (k)=\tilde{H}_{2}^{(1)} (k)+H_{2}^{(2)} (k)-k\delta \sigma ^{z} /2$, $\tilde{H}_{2}^{(1)} (k)=(-1)^{k+1} (\Omega _{k} /2)(\sigma ^{+} e^{-ik(\psi -\pi /2)} +h.c.)$, $\rho _{3}^{(k)} =U_{3}^{\dagger} \rho _{2} U_{3} $, $U_{3} =e^{-ik\delta \sigma ^{z} /2} $. The density matrix in the interaction representation in the laboratory frame is written as follows: $\rho ^{(k)} (t)=U_{1} U_{2} U_{3} e^{-i\tilde{H}_{eff} (k)t} U_{1}^{\dagger} \rho(0) U_{1} e^{i\tilde{H}_{eff} (k)t} U_{3}^{\dagger} U_{2}^{\dagger} U_{1}^{\dagger} $. We assume that the system is initially in the ground state ${\left| 0 \right\rangle} $. Using the equation for $\rho ^{(k)} (t)$, we obtain the probability to find the system in some moment again in the ground state $P_{{\left| 0 \right\rangle} }^{(k)} (t)={\left\langle 0 \right|} \rho ^{(k)} (t){\left| 0 \right\rangle} $:

\[P_{{\left| 0 \right\rangle} }^{(k)} (t)=\frac{1}{2} +\frac{1}{2} \cos ^{2} \theta -\left(\frac{\Omega _{k} }{\Omega _{k}^{*} } \right)^{2} \cos ^{2} \theta \sin ^{2} \frac{\Omega _{k}^{*} }{2} t+(-1)^{k+1} \frac{\Omega _{k} }{4\Omega _{k}^{*} } \sin \left(k(\psi -\frac{\pi }{2} )-a\cos \psi \right)\sin 2\theta \sin \Omega _{k}^{*} t-\]
\[-(-1)^{k+1} \frac{(\omega _{k}^{BS} -k\delta )\Omega _{k} }{2\Omega _{k}^{*2} } \cos \left(k(\psi -\frac{\pi }{2} )-a\cos \psi \right)\sin 2\theta \sin ^{2} \frac{\Omega _{k}^{*} }{2} t+\]
\[+\frac{1}{2} \sin \theta \left[-\left((-1)^{k+1} \frac{\Omega _{k}^{} }{\Omega _{k}^{*} } \sin (k\psi -\frac{k\pi }{2} )\cos \theta +\frac{\omega _{k}^{BS} -k\delta }{\Omega _{k}^{*} } \sin \theta \sin (a\cos \psi )\right)\sin \Omega _{k}^{*} t+\right. \]
\[+\left(\frac{\Omega _{k}^{2} }{\Omega _{k}^{*2} } \cos (2k\psi -k\pi -a\cos \psi )\sin \theta -\left(\frac{\omega _{k}^{BS} -k\delta }{\Omega _{k}^{*} } \right)^{2} \sin \theta \cos (a\cos \psi )-\right. \]
\[\left. -(-1)^{k+1} \frac{2\Omega _{k} (\omega _{k}^{BS} -k\delta )}{\Omega _{k}^{*2} } \cos (k\psi -\frac{k\pi }{2} )\cos \theta \right)\sin ^{2} \frac{\Omega _{k}^{*} }{2} t+\]
\[\left. +\sin \theta \cos (a\cos \psi )\cos ^{2} \frac{\Omega _{k}^{*} }{2} t\right]\cos \left(k\omega t-a\cos (\omega t+\psi )\right)+\]
\[+\frac{1}{2} \sin \theta \left[-\left((-1)^{k+1} \frac{\Omega _{k}^{} }{\Omega _{k}^{*} } \cos (k\psi -\frac{k\pi }{2} )\cos \theta +\frac{\omega _{k}^{BS} -k\delta }{\Omega _{k}^{*} } \sin \theta \cos (a\cos \psi )\right)\sin \Omega _{k}^{*} t+\right. \]
\[
+\left(\frac{\Omega _{k}^{2} }{\Omega _{k}^{*2} } \sin (a\cos \psi -2k\psi +k\pi )\sin \theta +\left(\frac{\omega _{k}^{BS} -k\delta }{\Omega _{k}^{*} } \right)^{2} \sin \theta \sin (a\cos \psi )+\right. \]
\[\left. +(-1)^{k+1} \frac{2\Omega _{k} (\omega _{k}^{BS} -k\delta )}{\Omega _{k}^{*2} } \sin (k\psi -\frac{k\pi }{2} )\cos \theta \right)\sin ^{2} \frac{\Omega _{k}^{*} }{2} t-
\]
\begin{equation}\label{eq_6}
\left. -\sin \theta \sin (a\cos \psi )\cos ^{2} \frac{\Omega _{k}^{*} }{2} t\right]\sin \left(k\omega t-a\cos (\omega t+\psi )\right),
\end{equation}
where
\begin{equation} \label{eq_7} \Omega _{k}^{*} =\sqrt{\Omega _{k}^{2} +(\omega _{k}^{BS} -k\delta )^{2} } \end{equation}
is the Rabi frequency in the non-RWA which takes into account the Bloch-Siegert-like shift.
\end{widetext}

Thus, the state population $P_{{\left| 0 \right\rangle} }^{(k)} $of the qubit level ${\left| 0 \right\rangle} $ for the $k$-th order Raman transition oscillates slowly at the non-RWA Rabi frequency $\Omega _{k}^{*} $ (the third, fourth and fifth terms in Eq. \eqref{eq_6}) and quickly at the frequencies which are integer multiples of the frequency $\omega $ (the other terms in Eq. \eqref{eq_6}). The amplitudes of the fast oscillations are slowly changed with the frequency $\Omega _{k}^{*} $. The fast multiphoton oscillations is caused by the longitudinal interaction $\sim \sin (\omega t+\psi )\sigma ^{z} $ in the Hamiltonian $H_{1} $. The non-RWA frequency $\Omega _{k}^{*} $ replaces the standard Rabi frequency, yielding a more precise value for the oscillation frequency of $P_{{\left| 0 \right\rangle} }^{(k)} $. Eq. \eqref{eq_7} shows that even at the strong driving the non-RWA frequency $\Omega _{k}^{*} $ can be equaled to the standard RWA Rabi frequency $\Omega _{k}$, when the Bloch-Siegert shift is compensated by the corresponding detuning $\delta$ (see section III B).

\subsection{Effective strength of Raman transitions}

Fig. 1 depicts the oscillations of the ground state population of the driven two-level spin system. The oscillations were calculated from Eq. \eqref{eq_6} with the Raman transitions with \textit{k} = 1, 2, 3, 4 assuming that the phase of low-frequency field $\psi =0$. The slow oscillations (red lines) at the non-RWA frequency $\Omega _{k}^{*} $ are accompanied by the fast oscillations  (green lines) at the frequencies which are integer multiples of the frequency $\omega $. The values of $\Omega _{k}^{*} $ characterize the effective strength of the spin-field coupling for Raman transitions and strongly decrease with increasing $k$.
\begin{figure}[t]
\centering\includegraphics[]{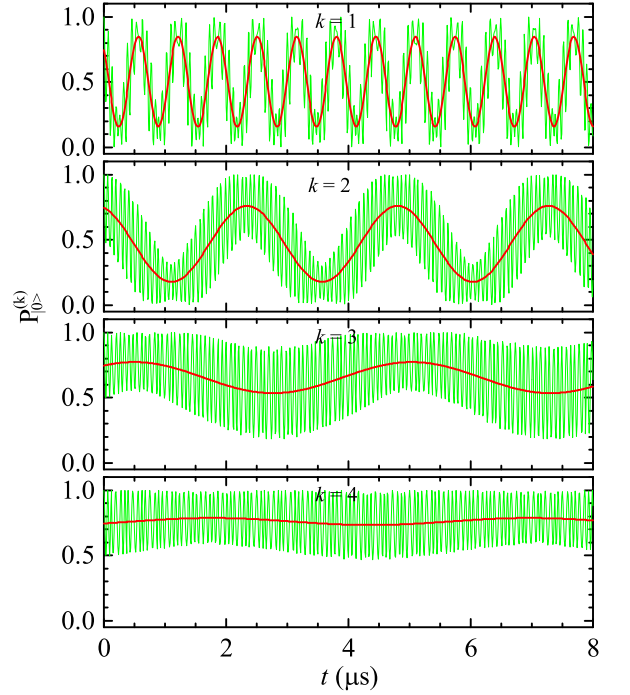}
\caption{The state population of the spin level ${\left| 0 \right\rangle} $ as a function of the evolution time for different Raman transitions. The strength of the low-frequency field is $A/2\pi =2. 22$ MHz and its frequency $\omega =\omega _{0} /k$, $\omega _{0} /2\pi =14. 17$ MHz, $\Delta _{x} /2\pi =10. 12$ MHz, $\Delta _{z} /2\pi =9. 92$ MHz, and $\psi =0$. The slow oscillations at the non-RWA frequency and the fast oscillations at the frequencies which are integer multiples of the frequency $\omega$ are shown by red and green lines, respectively. The system is initially in the ground state ${\left| 0 \right\rangle}$.}
\label{fig1}
\end{figure}
\subsection{Bloch-Siegert-like shift}

\begin{figure}[]
\centering
\includegraphics[]{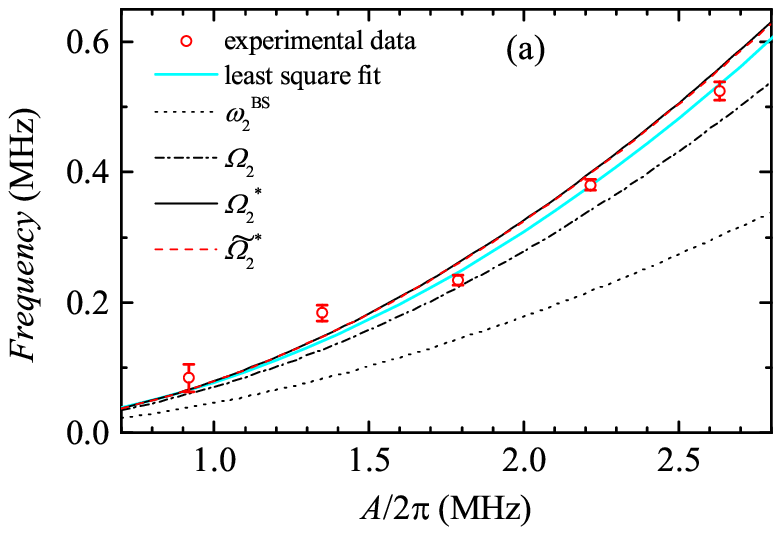}
\includegraphics[]{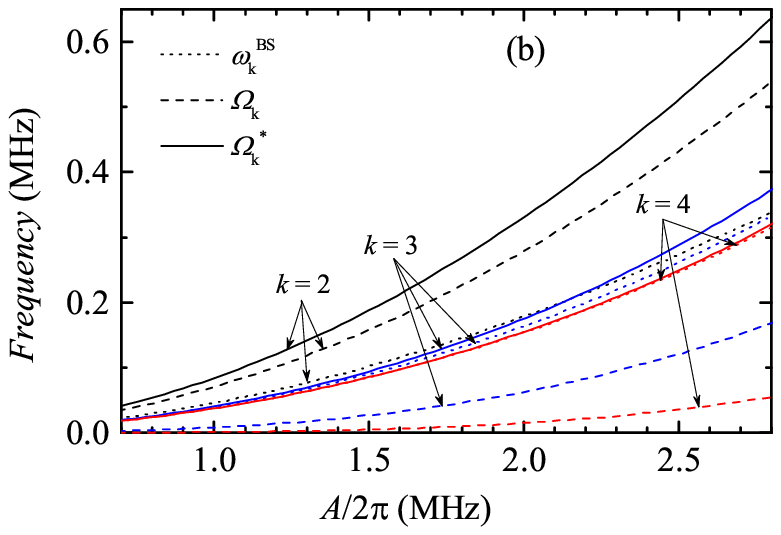}
\caption{The dependence of the RWA Rabi frequency $\Omega {}_{k} $, the non-RWA Rabi frequency $\Omega _{k}^{*} $ (and $\tilde{\Omega }_{2}^{*} $) and the Bloch-Siegert-like shift $\omega _{k}^{BS} $ on the amplitude of the driving field for different Raman transitions. The parametrs are $\omega =\omega _{0} /k+\delta $, $\omega _{0} /2\pi =14. 17$ MHz, $\Delta _{x} /2\pi =10. 12$ MHz, $\Delta _{z} /2\pi =9. 92$ MHz, and $\psi =0$. (a) The second order Raman transition. $\delta /2\pi =0. 005$ MHz. The open circles show the experimental data from Ref. \cite{pp24}. (b) The second, third and fourth order Raman transitions. $\delta=0$. }
\label{fig2}
\end{figure}
Fig. 2(a) shows the dependences of the values of the non-RWA Rabi frequency $\Omega _{2}^{*} $, the RWA Rabi frequency $\Omega _{2} $ and the Bloch-Siegert-like shift $\omega _{2}^{BS} $ on the strength of the low-frequency driving. They were calculated from Eqs. \eqref{eq_5} and \eqref{eq_7} for the second-order Raman transition ($\omega _{0} =2(\omega -\delta )$). The parameters of the driving field are the same as those used in the experiment \cite{pp24}. We also compare our calculations with the experimental data from Ref. \cite{pp24} and find that the calculated dependence of $\Omega _{2}^{*} (A)$ well approximates these data. One can see that it is impossible to obtain the quantitative agreement between the theoretical and experimental results using only $\Omega _{2} (A)$ without taking into account the Bloch-Siegert shift. At such driving the Bloch-Siegert shift is comparable with the value of the RWA Rabi frequency $\Omega _{2} $ value. The third-order correction $\Delta \Omega _{2} $ in $\Omega _{2}$ is presented in Appendix (see formula (A2)). The dependence $\tilde{\Omega }_{2}^{*} (A)$ with the third-order correction given by Eq. (A3) is shown in Fig. 2(a) by the dashed line. The effect of such correction is small for the used values of $A$ and $\omega $. The non-RWA Rabi frequencies for the second, third and fourth order Raman transitions are compared in Fig. 2(b). Unlike the second order transition, for the Raman transition with $k=3$ and $k=4$ the Bloch-Siegert frequency shift gives the dominant contribution to the non-RWA Rabi frequencies. The presented results obviously demonstrate that the driving field applied in the experiment \cite{pp24} cannot be considered as the weak one, because $A/\omega $ is about 0.3. Consequently, the obtained experimental results must be described beyond the RWA taking into account the antiresonant terms in the Hamiltonian \eqref{eq_2}. In fact, such description was realized in Ref. \cite{pp24} by numerical calculations. Note that the strong driving regime can easily be achieved for the dressed spin transitions. It is much more difficult\textit{ }to\textit{ }realize such regime for the bare spin transitions excited by the monochromatic field, because the driving strength is usually much weaker than the spin resonant frequency. In this case the strong driving of superconducting artificial atoms is commonly studied \cite{pp28}. Note that the curve for the non-RWA Rabi frequency calculated with the third-order correction is still outside of the error bars of the experimental points (Fig. 2 (a)). Because the least square fit of the results is also outside of the error bars of two experimental points, it is difficult to expect that higher order corrections would make the curve lay inside the error bars. Additional studies, including new experiments, are needed to find a reason of this discrepancy.

The obtained dependence of the non-RWA Rabi frequency $\Omega _{k}^{*} =\sqrt{\Omega _{k}^{2} +(\omega _{k}^{BS} -k\delta )^{2} } $ on detuning $\delta =\omega -\omega _{0} /k$ allows us to find directly the Rabi frequency in the RWA $\Omega _{k} $ for the \textit{k}-th Raman transition and the corresponding Bloch-Siegert-like shift using the position of maximum of the function $\Omega _{k}^{*} (\delta )$ (Fig. 3). The minimum frequency corresponding to the RWA Rabi frequency is obtained when the Bloch-Siegert-like shift is compensated by the positive value of $\delta $.

\begin{figure}[]
\centering
\includegraphics[]{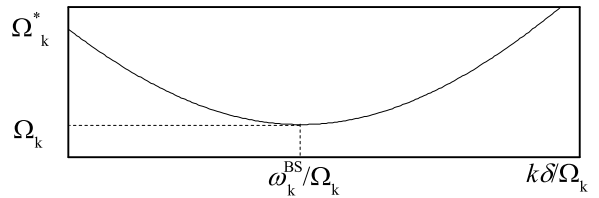}
\caption{The dependence of the non-RWA Rabi frequency $\Omega _{k}^{*} $ on detuning $\delta =\omega -\omega _{0} /k$ for the \textit{k}-th Raman transition. }
\label{fig3}
\end{figure}

\begin{figure}[]
\centering
\includegraphics[]{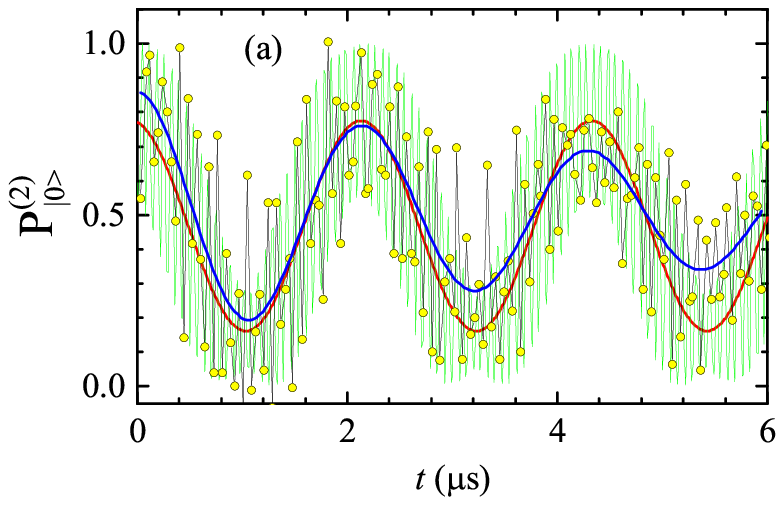}
\includegraphics[]{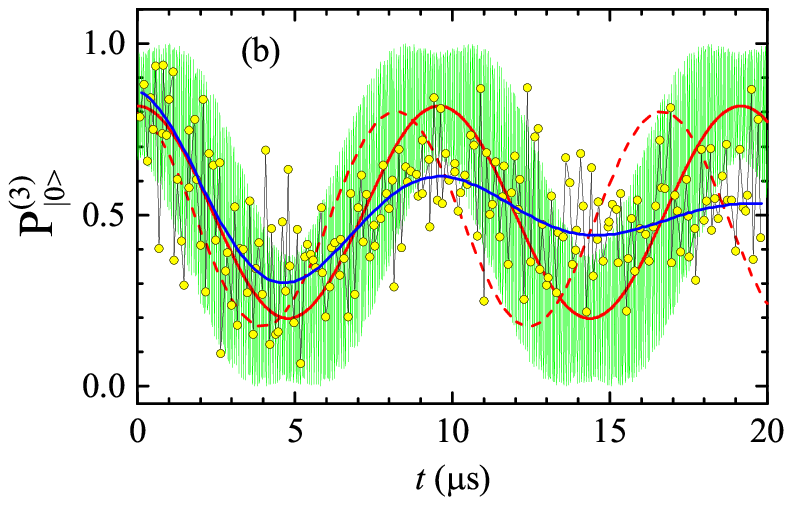}
\caption{The state population of the spin level ${\left| 0 \right\rangle} $ for the second and third order Raman transitions as a function of the evolution time. The decaying Rabi oscillations (blue lines) are a fit to the experimental data from \cite{pp24} presented by the open circles. The un-damped sinusoids (red lines) show our calculations of the Rabi oscillations accompanied with the fast oscillations. The strength and phase of the driving field are $A/2\pi =2. 37$ MHz and $\psi =0$. (a) The second order Raman transition at $\omega /2\pi =6. 985$ MHz, $\Delta _{x} /2\pi =9. 67$ MHz, $\Delta _{z} /2\pi =10. 03$ MHz, and $2\delta /2\pi =0. 038$ MHz. (b) The third order Raman transition at $\Delta _{x} /2\pi =9. 67$ MHz and $\Delta _{z} /2\pi =9. 82$ MHz. The red solid and dashed lines are the Rabi oscillations calculated at $3\delta /2\pi =0. 24$ MHz, $\omega /2\pi =4. 674$ and $3\delta /2\pi =0. 19$ MHz, $\omega /2\pi = 4.657$ MHz, respectively. The system is initially in the ground state ${\left| 0 \right\rangle}$.}
\label{fig4}
\end{figure}

Fig. 4 shows the state population of the spin level $|0\rangle$ as a function of the evolution time for the second- and third-order Raman transitions calculated from Eq. \eqref{eq_6} with the parameters used in the experiment \cite{pp24}. For the second order Raman transition there is a good agreement between the theoretical and experimental results (Fig. 4(a)). In this case the Bloch-Siegert-like shift $\omega _{2}^{BS} /2\pi =0. 27$ MHz is 7 times larger than the detuning $2\delta /2\pi =0. 038$ MHz. In accordance to Eq. \eqref{eq_7}, the observed oscillation is characterized by the non-RWA Rabi frequency $\Omega _{2}^{*}/2\pi =0. 455$ MHz, which differs from the RWA Rabi frequency $\Omega _{2}/2\pi =0. 394$ MHz. Another situation is realized for the third order Raman transition (Fig. 4(b)). At the detuning $3\delta /2\pi =0. 19$ MHz, given in Ref. \cite{pp24}, the calculated frequency of the Rabi oscillation (the red dashed  line) is something larger than the measured one. However, an increase of the detuning by 0. 05 MHz up to the value equaled to $\omega _{3}^{BS} /2\pi =0. 24$ MHz compensates the Bloch-Siegert effect and the oscillation (the red solid  line) occurs at the RWA Rabi frequency coinciding with the observed one.

\subsection{Effects of the initial phase of the low frequency field}

Features of the Rabi oscillations for Raman transitions strongly depend on the initial phase of the low-frequency field (Fig. 5). An increase of the phase of the driving field from $0$ to $90^{0} $decreases the amplitude of the Rabi oscillations and changes their phase. The case of the random phase, when the phase is stochastic and uniformly distributed from $0$ to $2\pi $, is also presented. The detailed phase dependences are presented in Fig. 6 for two amplitudes of the driving field. The amplitude and the phase of the Rabi oscillation for each $k$ shows periodic changes with a period $2\pi $. These changes weakly depend on the driving strength and are most visible for $k=3$ and $k=4$. The dependence of $P_{{\left| 0 \right\rangle} }^{(k)} (t)$ on the phase of the driving field gives additional possibilities for coherent control of a quantum system and can be used in dressed state engineering.

\begin{figure}[t]
\centering
\includegraphics[]{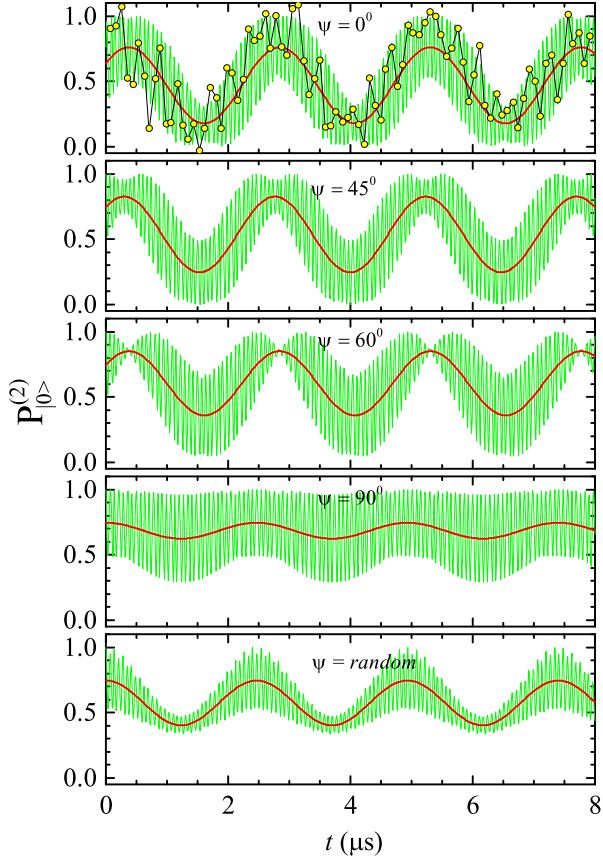}
\caption{The state population of the spin level ${\left| 0 \right\rangle} $ for the second order Raman transition as a function of the evolution time at different values of the phase $\psi $ of low-frequency field. The probability was calculated from Eq. \eqref{eq_6}. The open circles at $\psi = 0$ show the experimental data from \cite{pp24}. The strength and frequency of the driving field are $A/2\pi =2. 22$ MHz and $\omega /2\pi =7. 09$ MHz, respectively. The other parameters are $\Delta _{x} /2\pi =10. 12$ MHz, $\Delta _{z} /2\pi =9. 92$ MHz and $\delta /2\pi =0. 005$ MHz. The system is initially in the ground state ${\left| 0 \right\rangle}$.}
\label{fig5}
\end{figure}

\begin{figure}[t]
\includegraphics[]{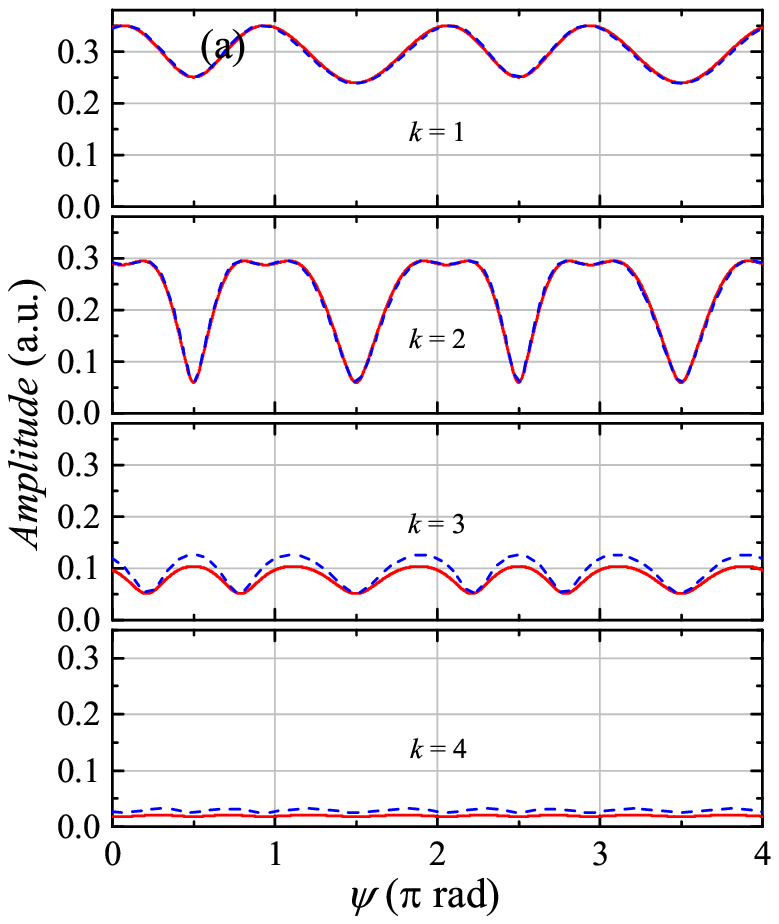}
\includegraphics[]{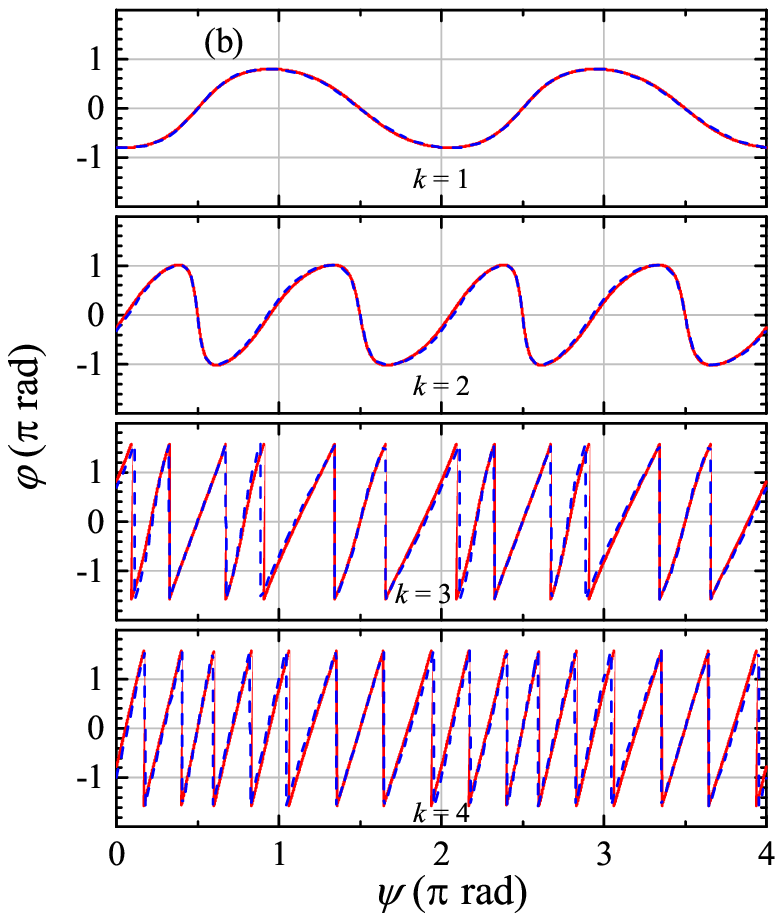}
\caption{The amplitude (a) and phase (b) of Rabi oscillation for different order Raman transitions versus the phase $\psi $ of the low-frequency driving field. The strength of the driving field is $A/2\pi =1. 79$ MHz (solid lines) and $A/2\pi =2. 22$ MHz (dashed lines); $\omega =\omega _{0} /k$, $\omega _{0} /2\pi =14. 17$ MHz, $\Delta _{x} /2\pi =10. 12$ MHz, and $\Delta _{z} /2\pi =9. 92$ MHz. }
\label{fig6}
\end{figure}

\subsection{Amplitudes of Rabi oscillations for multiphoton resonances versus detuning}

The dependences of the amplitudes of the Rabi oscillations on detuning $\delta =\omega -\omega _{0} /k$ for different Raman transitions are presented in Fig. 7. Due to the Bloch-Siegert effect the resonant frequencies are shifted from $\omega _{0} /k$. As mentioned above, the Bloch-Siegert shift can be compensated by the positive value of detuning $\delta $. This value does not exceed the half-width of the resonance lines presented in Fig. 7. At such detuning the non-RWA Rabi frequency $\Omega _{k}^{*} $ coincides with the RWA Rabi frequency $\Omega _{k} $ (see Eq. (7)). The multiphoton resonances become sharper with increasing the order of the Raman transition. To obtain the intensive Rabi oscillation, the resonant condition must be fulfilled more precisely for the higher order Raman transitions.

\begin{figure}[t]
\centering
\includegraphics[]{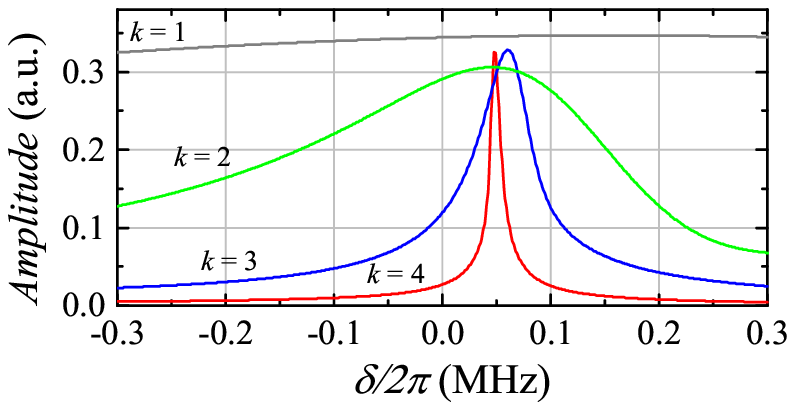}
\caption{The amplitude of Rabi oscillation for different Raman transitions versus detuning $\delta =\omega -\omega _{0} /k$. $A/2\pi =2. 22$ MHz, $\Delta _{x} /2\pi =10. 12$ MHz, $\Delta _{z} /2\pi =9. 92$ MHz, and $\omega _{0} /2\pi =14. 17$ MHz. }
\label{fig7}
\end{figure}

The presented close-form expressions and their applications for describing the coherent dynamics of Raman transitions in the driven solid-state spin system of NV center in diamond demonstrate that the Bloch-Siegert frequency shift gives a significant contribution to the frequency of multiphoton Rabi oscillations for $k=2$ and becomes dominant for $k=3$ and $k=4$. That is because the value of the coupling between the low-frequency driving field and the spin system is compared to the driving frequency. At such strong interaction the non-resonant processes of emission and absorption of photons become significant. As a result, the dressed states of the spin system are considerably shifted and the non-resonant processes essentially determine the values of multiphoton Rabi frequencies.

\section{CONCLUSIONS}

We have studied the coherent dynamics of multiphoton Raman transitions between dressed states in a two-level solid-state spin system driven by an amplitude-modulated microwave field. In the framework of the non-secular perturbation theory based on the Bogoliubov averaging method, closed-form expressions for the Rabi frequencies of these transitions have been obtained beyond the RWA for the low-frequency driving component. We have found the contribution of the Bloch-Siegert-like shift to the Rabi frequency and shown that for the high order Raman transitions this shift can dominate. We predicted the strong dependence of the amplitude and phase of the Rabi oscillations on the initial phase of the modulation field as well as on detuning from multiphoton resonance. The obtained results are in good agreement with the recent experimental data for nitrogen-vacancy center in diamond \cite{pp24}. The demonstrated multiphoton dynamics extends coherent quantum control in dressed state engineering by using the strong low-frequency driving.

\section{Acknowledgements}

The work was supported by Belarusian Republican Foundation for Fundamental Research (Grant F18R-257) and by State Programm of Scientific Investigations ``Physical material science, new materials and technologies'', 2016-2020.

\appendix{APPENDIX A}

For $k = 2$, the contribution of the third order in the parameter $A\cos \theta /\omega $ to the effective Hamiltonian of the non-secular perturbation theory based on the Bogoliubov averaging method is written as (see \cite{pp26,pp27})
\begin{widetext}
\[H_{2}^{(3)} =-\frac{1}{3} <[\int _{}^{t}d\tau (H_{2} (\tau )-<H_{2} (\tau )>) ,[\int _{}^{t}d\tau (H_{2} (\tau )-<H_{2} (\tau )>) ,(H_{2} (t)+\frac{1}{2} <H_{2} (t)>)]]>. \]
Then, we obtain
\[H_{2}^{(3)} =\frac{\Delta \Omega _{2} }{2} (\sigma ^{+} e^{-i2\psi } +h. c. ),\eqno(1A)\]
where
\[\Delta \Omega _{2} =\frac{A^{3} \cos ^{3} \theta }{6\omega ^{2} } \left\{\sum _{\begin{array}{l} {n\ne -3} \\ {m\ne -3} \end{array}}\frac{J_{n} J_{m} }{(n+3)(m+3)} \left(J_{n+m+3} +J_{n+m+5} +J_{-n+m-3} +J_{-n+m-1} \right)+ \right. \eqno(2A)\]
\[+\sum _{\begin{array}{l} {n\ne -3} \\ {m\ne -1} \end{array}}\frac{J_{n} J_{m} }{(n+3)(m+1)} \left(J_{n+m+1} +J_{n+m+3} +J_{-n+m-5} +J_{-n+m-3} \right) +\]
\[+\sum _{\begin{array}{l} {n\ne -1} \\ {m\ne -3} \end{array}}\frac{J_{n} J_{m} }{(n+1)(m+3)} \left(J_{n+m+1} +J_{n+m+3} +J_{-n+m-1} +J_{-n+m+1} \right) +\]
\[+\sum _{\begin{array}{l} {n\ne -1} \\ {m\ne -1} \end{array}}\frac{J_{n} J_{m} }{(n+1)(m+1)} \left(J_{n+m-1} +J_{n+m+1} +J_{-n+m-3} +J_{-n+m-1} \right)-\]
 \[\Biggl. -\frac{2J_{2} }{a} \sum _{n\ne -3}\frac{J_{n} }{(n+3)^{2} } \left(J_{n} +J_{n+2} -J_{-n-4} -J_{-n-6} \right) -\frac{2J_{2} }{a} \sum _{n\ne -1}\frac{J_{n} }{(n+1)^{2} } \left(J_{n} +J_{n-2} -J_{-n-4} -J_{-n-2} \right) \Biggr\},\]
\end{widetext}
and $\Delta \Omega _{2} $ is the third-order correction in $\Omega _{2} $. To simplify a notation of the formula, we omit the argument $a$ of the Bessel functions $J_{k} $. The non-RWA Rabi frequency with the third-order correction is rewritten as
\[\tilde{\Omega }_{2}^{*} =\sqrt{(\Omega _{2} +\Delta \Omega _{2} )^{2} +(\omega _{2}^{BS} -k\delta )^{2} }. \eqno(3A) \]
Since the third order of the perturbation theory gives the small correction, the contribution of $H_{2}^{(3)} (k)$ can be neglected. Then, the non-RWA Rabi frequency of the $k$-th order Raman transition can be expressed in the following approximate form:
\[\Omega _{k}^{*} \approx \sqrt{\Omega _{k}^{2} +(\omega _{k}^{BS} -k\delta )^{2} }. \eqno(4A)\]


\begin{thebibliography}{30}

\bibitem{pp1} \item T. Takui, L. Berliner, and G. Hanson, \textit{Electron spin resonance (ESR) based quantum computing}, (Springer, New York, 2016).

\bibitem{pp2} L. Allen and J. H. Eberly, \textit{Optical Resonance and Two-Level Atoms }(Wiley, New York, 1982).

\bibitem{pp3} J. R. Schaibley, A. P. Burgers, G. A. McCracken, D. G. Steel, A. S. Bracker, D. Gammon, and L. J. Sham, Phys. Rev. B \textbf{87}, 115311 (2013).

\bibitem{pp4} S. Shevchenko, S. Ashhab, and F. Nori, Physics Reports \textbf{492}, 1 (2010).

\bibitem{pp5} M. A. Nielsen and I. L. Chuang, \emph{Quantum Computation and Quantum Information (2nd ed. )}, (UniversityPress, Cambridge, 2010).

\bibitem{pp6} C. L. Degen, F. Reinhard, and P. Cappellaro,  Rev. Mod. Phys. \textbf{89}, 035002 (2017).

\bibitem{pp7} N. H. Lindner, G. Refael, and V. Galitski, Nature Physics \textbf{7}, 490 (2011).

\bibitem{pp8} S. Choi, J. Choi, R. Landig, G. Kucsko, H. Zhou, J. Isoya, F. Jelezko, S. Onoda, H. Sumiya, V. Khemani, C. von Keyserlingk, N. Y. Yao, E. Demler, and M. D. Lukin, Nature \textbf{543}, 221 (2017).

\bibitem{pp9} J. Zhang, P. W. Hess, A. Kyprianidis, P. Becker, A. Lee, J. Smith, G. Pagano, I. -D. Potirniche, A. C. Potter, A. Vishwanath, N. Y. Yao, and C. Monroe, Nature \textbf{543}, 201 (2017).

\bibitem{pp10} C. Cohen-Tannoudji, J. Dupont-Roc and G. Grynberg, \emph{Atom-Photon Interactions: Basic Processes and Applications} (John Wiley \& Sons, 1992).

\bibitem{pp11} S. N. Shevchenko, G. Oelsner, Ya. S. Greenberg, P. Macha, D. S. Karpov, M. Grajcar, U. H\"{u}bner, A. N. Omelyanchouk, and E. Il'ichev, Phys. Rev. B \textbf{89}, 184504 (2014).

\bibitem{pp12} G. Jeschke, Chem. Phys. Lett. \textbf{301}, 524 (1999).

\bibitem{pp13} A. P. Saiko and G. G. Fedoruk, JETP Letters \textbf{87}, 128 (2008).

\bibitem{pp14} S. Rohr, E. Dupont-Ferrier, B. Pigeau, P. Verlot, V. Jacques, and O. Arcizet, Phys. Rev. Lett. \textbf{112}, 010502 (2014).

\bibitem{pp15} A. Redfield, Phys. Rev. \textbf{98}, 1787 (1955).

\bibitem{pp16} R. Glenn, M. E. Limes, B. Pankovich, B. Saam, and M. E. Raikh, Phys. Rev. B \textbf{87}, 155128 (2013).

\bibitem{pp17} K. J. Layton, B. Tahayori, I. M. Y. Mareels, P. M. Farrell, and L. A. Johnston, J. Magn. Reson. \textbf{242}, 136 (2014).

\bibitem{pp18} H. Hatanaka, N. Tabuchi, J. Magn. Reson. \textbf{155}, 119 (2002).

\bibitem{pp19} S. Papademetriou, S. Chakmakjian, and J. C. R. Stroud, J. Opt. Soc. Am. B \textbf{9} (7), 1182 (1992).

\bibitem{pp20} Q. Wu, D. J. Gauthier, and T. W. Mossberg, Phys. Rev. A \textbf{50}, 1474 (1994).

\bibitem{pp21} C. C. Yu, J. R. Bochinski, T. M. V. Kordich, T. W. Mossberg, and Z. Ficek, Phys. Rev. A \textbf{56} R4381 (1997).

\bibitem{pp22} A. P. Saiko, R. Fedaruk, and S. A. Markevich, J. Magn. Reson. \textbf{259}, 47 (2015).

\bibitem{pp23} A. P. Saiko, R. Fedaruk, and S. A. Markevich, J. Phys. B \textbf{47}, 155502 (2014).

\bibitem{pp29} Y. Yan, Z. Lu, J.Y. Luo, and H. Zheng, Phys. Rev. A \textbf{96}, 033802 (2017).

\bibitem{pp30} Y. Yan, Z. Lu, J.Y. Luo, and H. Zheng, Phys. Rev. A \textbf{97}, 033817 (2018).

\bibitem{pp24} Z. Shu, Y. Liu, Q. Cao, P. Yang, S. Zhang, M. B. Plenio, F. Jelezko, and J. Cai. , arXiv:1804. 10492

\bibitem{pp25} A. Russomanno and G. E. Santoro, J. Stat. Mech. \textbf{2017}, 103104 (2017).

\bibitem{pp26} N. N. Bogoliubov and Yu. A. Mitropolsky. \emph{Asymptotic Methods in the Theory of Nonlinear Oscillations}, (Gordon and Breach, New York, 1961).

\bibitem{pp27} A. P. Saiko, S. A. Markevich, and R. Fedaruk, Phys. Rev. A \textbf{93}, 063834 (2016).

\bibitem{pp28} C. Deng, J. -L. Orgiazzi, F. Shen, S. Ashhab, and A. Lupascu, Phys. Rev. Lett. 115, 133601 (2015).

\end{thebibliography}
\end{document}